\documentclass[twocolumn,tighten]{aastex62}
\usepackage{amssymb, amsmath,framed}

\usepackage{bm}
\expandafter\ifx\csname package@font\endcsname\relax\else
 \expandafter\expandafter
 \expandafter\usepackage
 \expandafter\expandafter
 \expandafter{\csname package@font\endcsname}
\fi
\def\bq{\begin{equation}}
\def\eq{\end{equation}}
\def\bqy{\begin{eqnarray}}
\def\eqy{\end{eqnarray}}

\begin{document}
\title{\Large{Optimal Target Stars in the Search for Life}}

\correspondingauthor{Manasvi Lingam}
\email{manasvi.lingam@cfa.harvard.edu}

\author{Manasvi Lingam}
\affiliation{Institute for Theory and Computation, Harvard University, Cambridge, MA 02138, USA}
\affiliation{Harvard-Smithsonian Center for Astrophysics, Cambridge, MA 02138, USA}

\author{Abraham Loeb}
\affiliation{Institute for Theory and Computation, Harvard University, Cambridge, MA 02138, USA}
\affiliation{Harvard-Smithsonian Center for Astrophysics, Cambridge, MA 02138, USA}

\begin{abstract}
The selection of optimal targets in the search for life represents a highly important strategic issue. In this Letter, we evaluate the benefits of searching for life around a potentially habitable planet orbiting a star of arbitrary mass relative to a similar planet around a Sun-like star. If recent physical arguments implying that the habitability of planets orbiting low-mass stars is selectively suppressed are correct, we find that planets around solar-type stars may represent the optimal targets.\\
\end{abstract}

\section{Introduction}
Recently, several papers have emphasized the need for caution when scientists convey the latest discoveries of potentially habitable exoplanets, as habitability metrics and terminology are readily susceptible to misinterpretation \citep{TT17,ML17}.\footnote{For instance, there has been a tendency in some quarters to associate the concept of the habitable zone of the host star with the broader notion of habitability.} As a result, there is a real possibility that the combination of misleading terminology and the non-detection of the fingerprints of life could lead to public disenchantment, and subsequent cuts in related federal funding.

This unfortunate outcome emerged after several decades of NASA funding toward the Search for Extraterrestrial Intelligence (SETI), when the last major program was shut down in 1993 \citep{Garb99}. A significant factor responsible for the cancellation of SETI funding arguably stemmed from the fact that decades of searching had not yielded any results, despite the fact that the region of parameter space covered was extremely small.\footnote{In addition, the ``giggle factor'' associated with SETI also played an important role in halting federal funding \citep{WOR18}.} A different example concerns the announcement of microfossils in the Martian meteorite ALH84001 in 1996 \citep{MG96}. Although the majority of this meteorite's unique features have been argued to be non-biogenic in origin \citep{MY12}, the public interest and the scientific impetus imparted to astrobiology at that time were unprecedented.\footnote{\url{https://www2.jpl.nasa.gov/snc/clinton.html}} Even if the detection of these microfossils may have been a false positive of sorts, the upcoming \emph{ExoMars} mission, which will search for past life on Mars, suggests that the scientific and public interest in Mars' biological potential has clearly not come to an interminable halt.\footnote{\url{http://exploration.esa.int/mars/}}

The availability of funding comes down to the perceived credibility and successes of a given astrobiological endeavor. Hence, the identification of optimal search strategies for biosignatures will be helpful (arguably necessary) in enhancing the chances of success \citep{HJ10}, and thereby improving the credibility of such missions. It is the purpose of this Letter to examine this question, and pave the way toward the selection of optimal target stars in the search for life's signatures; we note that similar topics have been explored recently by \citet{BAK17} and \citet{KGO18}.

\section{Relative merits of searching for life around stars of different masses}
We will present our methodology for assessing the benefits of searching for life around stars with different masses, and subsequently discuss the implications. 

\subsection{Methodology}
Our approach is based on the standard cost-benefit analyses often used in economics \citep{LG94,Bre06}, where the potential benefits ($B$) are weighed against the costs ($C$). We consider two scenarios: the detection of biosignatures (or technosignatures) on an ``Earth-analog'' (a planet with basic physical parameters similar to Earth) orbiting (i) an arbitrary star with a general mass $M_\star$, and (ii) a G-type star of approximately solar mass $M_\odot$. We will denote all of the quantities associated with (i) and (ii) via the subscripts `$\star$' and `$\odot$' respectively. 

The total cost of the telescope ($C$) can be expressed as the sum of the capital expenditure and the operational expenditure over the telescope's lifetime. Because the same telescope would be employed for detecting biosignatures around different stars, we focus on the expected benefits. It is well known that the socioeconomic returns from basic scientific research are not readily quantifiable \citep{Pos04}. Nevertheless, we can estimate the total benefit ($B$) arising from this enterprise as
\begin{equation} \label{defB}
    B \propto P \cdot \mathcal{P},
\end{equation}
where $P$ denotes the probability that the chosen planet has life, and $\mathcal{P}$ is the probability that the biosignatures arising from this exolife are detectable.\footnote{Note that this formula is akin to estimating the economic risks posed by catastrophic events, where the cost is calculated from the product of the event's probability and the predicted economic damage \citep{Pos04}.} The proportionality constant in (\ref{defB}) signifies the total scientific benefit that would be derived from the discovery of exolife.

An ideal mission should attempt to maximize the value of $B/C$. In order to compare the \emph{relative} benefits of the strategies (i) and (ii), we introduce the variable 
\begin{equation}\label{Delta}
    \Delta = \frac{B_\star}{B_\odot} = \left(\frac{P_\star}{P_\odot}\right)  \left(\frac{\mathcal{P}_\star}{\mathcal{P}_\odot}\right),
\end{equation}
where the last equality follows from (\ref{defB}). If $\Delta < 1$, searching for life around solar-type stars is a more effective strategy while the converse is true for $\Delta > 1$.

\subsection{Probability of life on Earth-analogs around stars of different masses}

From (\ref{Delta}), it is clear that $\Delta$ depends on the ratio $\mu \equiv P_\star/P_\odot$, which quantifies the ratio of the probabilities of life-bearing planets around stars of mass $M_\star$ and $M_\odot$. The value of $\mu$, which can be interpreted as the relative likelihood of life, is currently unknown. There are several natural possibilities. \\

\noindent{\bf (1) Flat ``prior'' (FP):} It is conceivable that $P_\star$ is independent of $M_\star$, and is regulated solely by planetary parameters. This would imply that the probability of finding life-bearing planets is equal for all stars (up to some upper bound), and that $\mu = 1$. \\

\noindent{\bf (2) Physical constraints on habitability (PH):} The habitability of planets orbiting M-dwarfs has famously swung back and forth over the past decade \citep{SBJ16}. Recent physical constraints on long-term habitability--for e.g. paucity of bioactive ultraviolet (UV) radiation, weak magnetic fields, compressed planetary magnetospheres, enhanced atmospheric erosion, rapid water loss, and frequent flares \citep{VJ13,LB15,GDC16,GDC17,MLin17,DHL17,DLMC,DJL18}--appear to indicate that life-bearing planets around M-dwarfs are likely to be rare relative to Sun-like stars \citep{Dole,KWR93,Kas10,HA14,Man17}. However, many of these potentially detrimental factors could have valid counterarguments. For the sake of simplicity, we focus solely on atmospheric escape driven by the stellar wind \citep{ZSR10}; other constraints (e.g. stellar UV radiation and lifetime) can be incorporated in a similar fashion \citep{LL17}. As the maximal duration over which evolution can occur is constrained by the escape timescale \citep{Man17}, which in turn is inversely proportional to the escape rate, we get
\begin{equation}\label{Atm}
    \mu \sim \left(\frac{L_\star}{L_\odot}\right) \left(\frac{M_\star}{M_\odot}\right)^{1.76},
\end{equation}
by using Equations (10) and (11) from \citet{LiLo17}. We have also made use of $a \propto L_\star^{1/2}$, $\Omega_\star \sim \Omega_\odot$, $R_p \sim R_\oplus$ and $P_s \sim 1$ atm for an Earth-analog. Note that $R_p$, $P_s$ and $a$ denote the planet's radius, surface pressure and semi-major axis respectively, while $L_\star$ and $\Omega_\star$ denote the stellar luminosity and rotation rate; $L_\star$ can be further expressed in terms of $M_\star$ via the mass-luminosity relationship \citep{LBS16}.\\

\noindent{\bf (3) Copernican Principle (CP):} It is also possible to apply the Copernican Principle (also called the Principle of Mediocrity) to obtain a heuristic estimate of $\mu$ as follows. Consider the following product:
\begin{equation}
    N_\star = n_\star \cdot \tau_\star \cdot f_\star \cdot P_\star \cdot P^{(I)}_\star,
\end{equation}
where $n_\star$ denotes the number of stars of mass $M_\star$, $\tau_\star$ is the total lifetime of the star, $f_\star$ represents the number of rocky planets in the habitable zone (HZ) of the host star, $P_\star$ is the probability that the planet supports life of some sort, and $P^{(I)}_\star$ is the probability that the life present on the planet is ``intelligent''.  Hence, $N_\star$ can be viewed as the total spatio-temporal likelihood of intelligent life existing on rocky planets (in the HZ) orbiting stars of mass $M_\star$. Consequently, the ratio $\delta = N_\star/N_\odot$ signifies the overall likelihood of intelligent life existing on rocky planets in the HZ of stars with mass $M_\star$ relative to rocky planets in the HZ around solar-analogs. 

Now, let us suppose that $\delta \gg 1$. In this case, the existence of a technologically intelligent species on the Earth orbiting a solar-type star in the current epoch would be anomalous. This would lead to a violation of the Copernican Principle, as the Earth would be rendered highly atypical. Hence, the choice $\delta \lesssim 1$ would be plausible for $M_\star \lesssim M_\odot$ (although the regime $\delta > 1$ cannot be ruled out). From Table 1 of \citet{Kal17}, it can be seen that $f_\star/f_\odot \sim 1$. We cannot evaluate the ratio $P^{(I)}_\star/P^{(I)}_\odot$, since the evolution of intelligence was ostensibly an important evolutionary innovation \citep{Car08,Wat08,LL18}, but only one such event in a series of megatrajectories \citep{KB00}. The extent to which evolutionary transitions (including technological intelligence) are universal on exoplanets is almost impossible to judge because of the interplay of contingent and convergent evolutionary processes, and the coupling between organisms and their environment \citep{LMan17}. We shall therefore assume henceforth that $P^{(I)}_\star/P^{(I)}_\odot \sim 1$, but see \citet{LL17} and \citet{LL18}. 

Using all of the above facts, we find that
\begin{equation} \label{muCP}
    \mu \sim \delta \left(\frac{\tau_\star}{\tau_\odot}\right)^{-1}  \left(\frac{n_\star}{n_\odot}\right)^{-1}.
\end{equation}
The second and third factors present in the RHS can be computed by using the empirical expressions provided in \citet{LBS16} and \citet{Kro01} respectively. Thus, $\delta \lesssim 1$ provides a heuristic upper bound for $\mu$ based on the Copernican Principle. Using (\ref{muCP}), we find that the likelihood of life around M-dwarfs is suppressed by several orders of magnitude relative to solar-type stars; this result is qualitatively consistent with similar statistical analyses \citep{Wal11}.

\begin{figure}
\includegraphics[width=7.5cm]{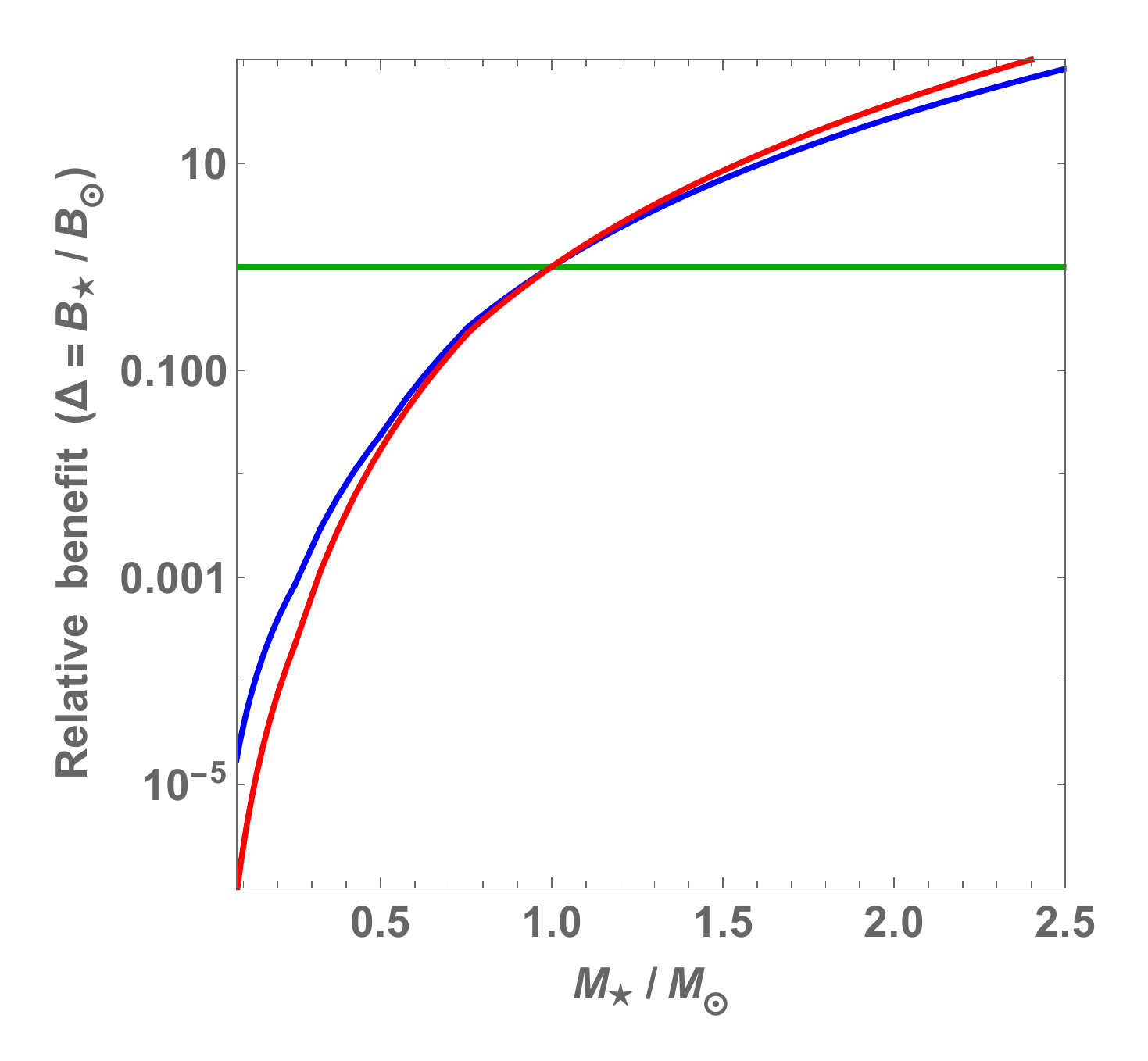} \\
\caption{Relative benefit $\Delta$ of an ideal search strategy as a function of stellar mass. The blue curve denotes the case where $\mu \equiv P_\star/P_\odot$ is given by (\ref{muCP}) with $\delta \approx 1$. The red curve represents the case where $\mu$ is governed by (\ref{Atm}), while the green curve corresponds to the flat prior of $\mu = 1$.}
\label{FigIdeal}
\end{figure}

\subsection{An ideal search strategy scenario}\label{SSecIdeal}

We commence our analysis of $\Delta$ by considering the highly idealized scenario wherein we possess telescopes with arbitrarily high resolution, accuracy, signal-to-noise ratio (S/N), etc. In this case, the detection of life is always guaranteed provided that it exists on the planet under consideration. Hence, it follows that $\mathcal{P}_\star = \mathcal{P}_\odot = 1$ and we obtain $\Delta = \mu$ upon using (\ref{Delta}). 

For the case FP, we see that $\Delta = 1$ implying that the search for life around M-dwarfs is likely to be equally beneficial compared to solar-analogs. In contrast, if we consider the cases PH (red curve) and CP (blue curve), the behavior of $\Delta$ is rendered very different as seen from Fig. \ref{FigIdeal}.\footnote{Since $\Delta = \mu$ for the scenario involving ideal detection capabilities, Fig. \ref{FigIdeal} also serves as the plot of the relative likelihood of life $\mu$, computed via (\ref{Atm}) and (\ref{muCP}).} Here, it must be noted that the blue curve may represent an upper bound on the relative scientific benefit $\Delta$. For cases PH and CP, we observe that $\Delta$ is reduced considerably, by several orders of magnitude, for a planet orbiting a low-mass star compared to one around a solar-analog. Hence, an Earth-analog in the HZ of a G-type star appears to merit a higher priority relative to one in the HZ of an M-dwarf in the context of searching for biosignatures. 

\subsection{Search strategies with observational constraints}\label{SSecObs}

\begin{figure}
\includegraphics[width=7.2cm]{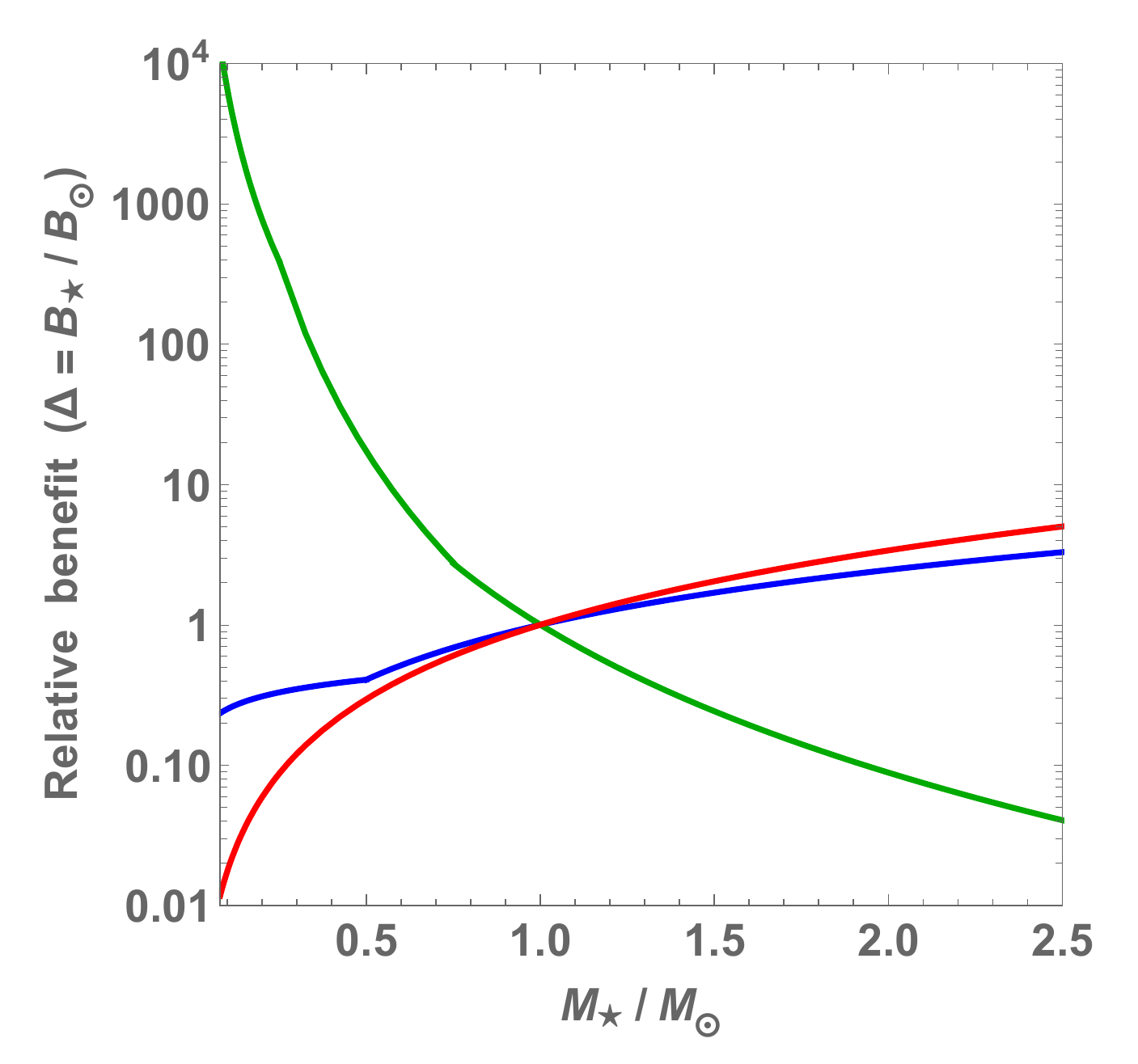} \\
\caption{Relative benefit $\Delta$ for a direct-imaging search strategy as a function of stellar mass. The blue curve denotes the case where $\mu$ is governed by (\ref{muCP}) with $\delta \approx 1$. The red curve represents the case with $\mu$ given by (\ref{Atm}), while the green curve corresponds to the flat prior of $\mu = 1$.}
\label{FigReal}
\end{figure}

Hitherto, we have considered an ideal scenario where $\mathcal{P}$ was independent of $M_\star$. The reality is manifestly different, because the ease of detecting biosignatures around habitable planets will vary significantly based on the star under consideration. Moreover, an important factor worth highlighting is that the method used for detecting biosignatures will influence the nature of $\kappa \equiv \mathcal{P}_\star/\mathcal{P}_\odot$, implying that the scaling for $\kappa$ is method dependent. Some of the widely considered strategies include high-contrast imaging, transmission and eclipse spectroscopy. 

For example, let us consider direct imaging as the search strategy. An important criterion for using direct imaging is that the inner working angle (IWA) must be smaller than the star-planet angular separation \citep{FA17}. Apart from this non-trivial requirement, the use of starshades or coronographs to suppress the starlight to the level of scattered planetary light is necessary. The near-IR contrast between the latter and the former, denoted by $\mathcal{C}$, is
\begin{equation} \label{Cont}
    \mathcal{C} \sim 10^{-10}\,\left(\frac{R_p}{R_\oplus}\right)^2 \left(\frac{a}{1\,\mathrm{AU}}\right)^{-2}\left(\frac{A}{0.3}\right),
\end{equation}
where $A$ is the planet's albedo. If the above two conditions are satisfied, the detection of both atmospheric and surface biosignatures become potentially feasible \citep{FA17}. Assuming that the probability of detection is proportional to the contrast $\mathcal{C}$, from (\ref{Cont}) and the preceding discussion, we find
\begin{equation}
    \kappa \sim \left(\frac{L_\star}{L_\odot}\right)^{-1},
\end{equation}
for an Earth-analog. By combining this expression with our estimates for $\mu$, the value of $\Delta$ can be calculated. We have plotted the two cases for $\Delta$ in Fig. \ref{FigReal}. The green curve, which corresponds to the case FP, clearly demonstrates that searching for biosignatures on an Earth-analog orbiting a lower-mass star is more efficient. The reason primarily stems from the greater ease of detecting biosignatures, since it has been assumed that all planets have an equal (i.e. mass-independent) probability of hosting biospheres. 

On the other hand, taking the case PH (red curve) into consideration implies that $\Delta < 1$ for $M < M_\odot$, indicating that searching for biosignatures around a low-mass star confers less benefits relative to a solar-type star. For the case CP (blue curve), we find that $\Delta$ becomes weakly dependent on $M_\star$. Here, the greater ease of detecting biosignatures is counteracted by the fact that the probability of life-bearing planets around low-mass stars is suppressed in accordance with (\ref{muCP}). In this scenario, we find that $\Delta \lesssim 1$ for $M_\star \lesssim M_\odot$, implying that it may be somewhat more efficient to conduct searches around solar-type stars. At this stage, an important observation is necessary: the case CP was computed assuming $\delta \sim 1$, and hence the blue curve constitutes an upper bound. The case $\delta \ll 1$ cannot be ruled out, thereby implying that $\delta$ can become arbitrarily small. In this regime, $\Delta \ll 1$ is possible for low-mass stars, bolstering the conclusion that low-mass stars do not represent efficient search targets when the case CP is considered. 

\begin{figure}
\includegraphics[width=7.2cm]{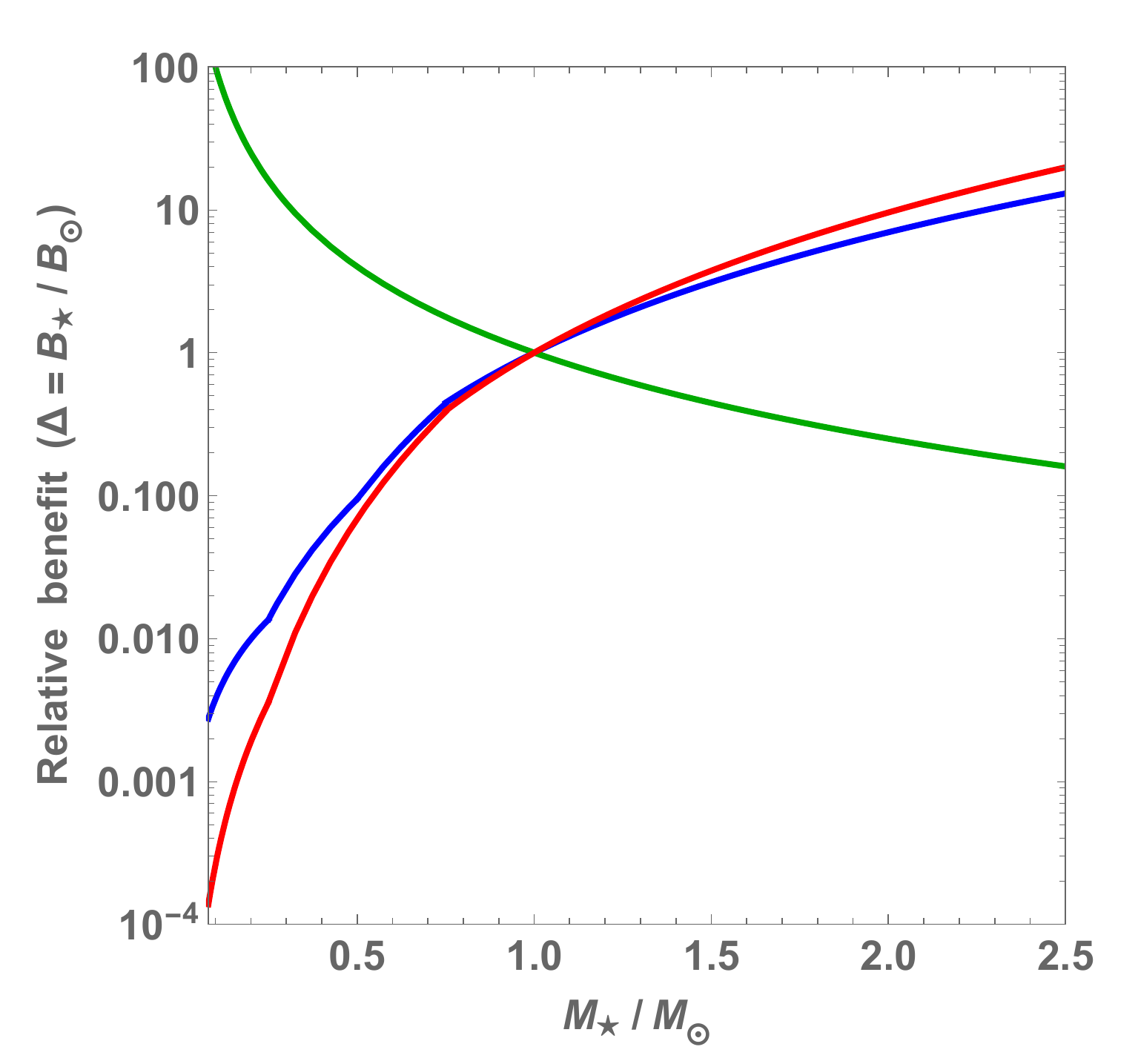} \\
\caption{Relative benefit $\Delta$ for a transmission spectroscopy search strategy as a function of stellar mass. The blue curve denotes the case where $\mu$ is governed by (\ref{muCP}) with $\delta \approx 1$. The red curve represents the case with $\mu$ given by (\ref{Atm}), while the green curve corresponds to the flat prior of $\mu = 1$.}
\label{FigRealT}
\end{figure}

Next, let us consider the use of transmission spectroscopy as the search strategy. Transmission spectroscopy is suitable for characterizing planetary atmospheres, and thereby potentially detecting biosignature gases such as oxygen, ozone and methane (especially if the telescope can operate in the UV range). The magnitude $S$ of these spectral features (relative to the stellar flux) is estimated \citep{FA17} via
\begin{equation}
    S \sim 84\,\mathrm{ppm}\,\left(\frac{N_H}{4}\right)\left(\frac{\mathcal{H}}{8\,\mathrm{km}}\right)\left(\frac{R_p}{R_\oplus}\right) \left(\frac{R_\star}{0.1\,R_\odot}\right)^{-2},
\end{equation}
where $R_\star$ is the stellar radius, $\mathcal{H}$ is the atmospheric scale height, and $N_H$ is a dimensionless constant that accounts for the depth of absorption features, and is dependent on the atmospheric composition. In addition, the S/N must be sufficiently high, and it can be calculated in the idealized limit where only photon noise is present \citep{FA17}. Assuming that this criterion is satisfied, for Earth-analogs we find that $S \propto M_\star^{-2}$ (because the planetary parameters drop out), where we have made use of the approximate mass-radius relationship $R_\star \propto M_\star$ \citep{DK91}. Thus, if $\kappa$ is proportional to $S$, we find
\begin{equation}
    \kappa \sim \left(\frac{M_\star}{M_\odot}\right)^{-2}.
\end{equation}
The resulting value of $\Delta$ has been plotted in Fig. \ref{FigRealT}. From the figure, we see that the case FP implies that searching for biosignatures around an Earth-analog orbiting an M-dwarf yields greater benefits. In contrast, if the probability is suppressed around low-mass stars (i.e. the cases PH and CP), we find that priority should be given to an Earth-analog around a solar-mass star. 

\subsection{A discussion of survey strategies}
Until now, we have focused on the scenarios (i) and (ii), wherein a terrestrial planet around a \emph{specific} star constitutes the target. We can now ask the same question, i.e. determining what constitutes the optimal search strategy, but for survey missions.

Let us suppose that the minimum photon flux detectable by a telescope is denoted by $\Phi$, and the maximum achievable survey distance by $d_\star$ (for a star with mass $M_\star$). As $\Phi \propto L_\star/d_\star^2$, we obtain $d_\star \propto L_\star^{1/2}$. If we assume that $d_\star \sin \theta < H_g$, where $\theta$ is the angular reach of the survey and $H_g$ is the Galactic scale height, the number of stars $\mathcal{N}_\star$ covered in the survey is $\mathcal{N}_\star \propto n_\star d_\star^3 \propto n_\star L_\star^{3/2}$. Thus, introducing the ratio $\eta \equiv  \mathcal{N}_\star/ \mathcal{N}_\odot$, we obtain
\begin{equation}
    \eta \sim \left(\frac{n_\star}{n_\odot}\right) \left(\frac{L_\star}{L_\odot}\right)^{3/2}.
\end{equation}
On the other hand, if $d_\star \sin \theta > H_g$, we have $ \mathcal{N}_\star \propto n_\star d_\star^2 H_g$, and this leads us to
\begin{equation}
    \eta \sim \left(\frac{n_\star}{n_\odot}\right) \left(\frac{L_\star}{L_\odot}\right). 
\end{equation}

To estimate the relative benefit of surveys of low-mass stars versus that of solar-type stars, the previous estimates for $\Delta$ must be multiplied with the factor $\eta$ introduced earlier. Upon doing so, we find that the preceding results are not greatly altered, and the key conclusions are delineated below:
\begin{itemize}
    \item For the idealized limit (with unlimited observational power) discussed in Sec. \ref{SSecIdeal}, we conclude that surveying solar analogs yields more benefits compared to surveys of M-dwarfs.
    \item In contrast, for the search strategies outlined in Sec. \ref{SSecObs}, we find that the benefits depend on the choice of prior. If we consider the case FP (flat prior), the survey of M-dwarfs may be more advantageous relative to solar-mass stars. On the other hand, if we consider the cases PH and CP (with habitability suppressed around M-dwarfs), the survey of Sun-like stars is more suitable for detecting life compared to surveying low-mass stars.
\end{itemize}

\section{Conclusions}
The key strategic challenge in the search for life is the selection of the target stars and planets. If the targets are poorly chosen and the search fails to detect signatures of life, there is a risk that the search for primitive (i.e. non-technological) forms of life will share the fate of SETI and lose its mainstream credibility and federal funding. Current surveys ``search for the keys under the lamp post'' by focusing on the easier to detect targets of planets in the HZ of M-dwarfs.

In this Letter, we have attempted to assess the relative benefits of searching for biosignatures on an Earth-analog in the HZ of a star with mass $M_\star$ relative to one orbiting a solar-type star. We have found that two broad conclusions can be drawn. First, if observational constraints arising from telescope sensitivity are not an issue, searches for life around solar-mass stars are more advantageous compared to M-dwarfs. Second, if certain observational constraints are taken into account, the choice of strategy is sensitive to the choice of prior for life-bearing planets as a function of the stellar mass.\footnote{This conclusion is analogous to the issue of whether abiogenesis on life-bearing planets is ``fast'' or ``slow'', wherein the choice of prior plays a dominant role \citep{ST12}.} 

If we consider a flat prior, where the probability of life is independent of the choice of star, focusing on planets around M-dwarfs is more advantageous because the detection of biosignatures becomes much easier. On the other hand, there is mounting evidence, especially based on considerations of space weather, that the likelihood of habitability for Earth-analogs around M-dwarfs might be much lower relative to their counterparts around G-type stars. Hence, if we adopt a prior where the habitability is selectively suppressed around low-mass stars, we conclude that it would be more advantageous to focus on the search for life on planets orbiting Sun-like stars relative to those around M-dwarfs.

It goes without saying that our analysis represents a simplified treatment. Apart from the mass $M_\star$, we have not taken stellar parameters into account, and focused exclusively on Earth-analogs. Similarly, our consideration of observational constraints and strategies is by no means exhaustive. Our work mostly relied on frequentist inferences, and more sophisticated treatments (using Bayesian frameworks) are worth undertaking.

Despite these caveats, we believe that there is a case for reassessing the optimal search strategies for extraterrestrial life, especially in light of the funding and credibility risks involved. Hence, it is our sincere hope that this work will stimulate further analyses of this topic that incorporate more detailed observational and theoretical constraints based on the rapidly burgeoning field of exoplanetary habitability.

\acknowledgments
This work was supported in part by grants from the Breakthrough Prize Foundation for the Starshot Initiative and Harvard University's Faculty of Arts and Sciences, and by the Institute for Theory and Computation (ITC) at Harvard University.


\end{document}